\documentclass{article}

\usepackage{arxiv}

\usepackage[utf8]{inputenc} 
\usepackage[T1]{fontenc}    
\usepackage{hyperref}       
\usepackage{url}            
\usepackage{booktabs}       
\usepackage{amsfonts}       
\usepackage{nicefrac}       
\usepackage{microtype}      
\usepackage{lipsum}
\usepackage{graphicx}
\usepackage{siunitx}
\usepackage{amsmath}

\graphicspath{ {./images/} }

\title{Scalable Motion In-betweening via Diffusion and Physics-Based Character Adaptation}

\author{
 Jia Qin \\
  State Key Laboratory of CAD\&CG \\
  Zhejiang University \\
  \texttt{qinjia@zju.edu.cn} \\
}

\begin{document}
\maketitle

\begin{abstract}
We propose a two-stage framework for motion in-betweening that combines diffusion-based motion generation with physics-based character adaptation. In Stage 1, a character-agnostic diffusion model synthesizes transitions from sparse keyframes on a canonical skeleton, allowing the same model to generalize across diverse characters. In Stage 2, a reinforcement learning-based controller adapts the canonical motion to the target character’s morphology and dynamics, correcting artifacts and enhancing stylistic realism. This design supports scalable motion generation across characters with diverse skeletons without retraining the entire model. Experiments on standard benchmarks and stylized characters demonstrate that our method produces physically plausible, style-consistent motions under sparse and long-range constraints.
\end{abstract}

\keywords{Motion In-betweening \and Diffusion Models \and Physics-based Character Animation}

\section{Introduction}

From animated films and video games to virtual avatars and digital twins, high-quality character animation plays a central role in modern digital media. 
One of the most common and time-consuming tasks in animation production is hand-crafting motion via keyframe animation, with in-between frames generated automatically.
Although motion capture technology offers an efficient way to acquire realistic human motion, it remains limited in several important aspects. 
First, captured data often requires extensive post-processing, such as denoising and cleanup. 
Second, actors may struggle to perform stylized or physically exaggerated motions that go beyond human capabilities. 
Third, even when successful, captured motions often need to be retargeted to characters with different skeletal structures or body proportions, introducing further complexity and potential artifacts.
These limitations restrict the direct applicability of motion capture in many production scenarios. As a result, manual keyframing and automatic in-betweening remain indispensable tools in modern animation workflows.

Recent advances in diffusion-based motion models have significantly improved the quality of in-betweening under sparse keyframe constraints. These models are capable of producing realistic and diverse motions by learning from large-scale human motion datasets. 
However, most existing methods are built on a fixed skeletal structure~\cite{tevet2022human, qin2024robust}, making adaptation to new characters challenging and often requiring retraining the entire model.
Additionally, many of these approaches operate purely in the kinematic domain~\cite{qin2022motion, oreshkin2023motion}, lacking physical awareness. As a result, when generating motion for stylized characters, 
they frequently produce artifacts such as foot sliding, ground penetration, and self-intersections, especially when kinematic retargeting is applied to skeletons with differing proportions.

To address these limitations, we propose a two-stage framework that combines the generative flexibility of diffusion models with the physical realism of simulation-based adaptation. In Stage 1, we use a diffusion model to generate motion sequences on a canonical skeleton from sparse keyframes. This stage is character-agnostic, enabling generalization across a wide range of characters. In Stage 2, we train a character-specific controller via reinforcement learning to adapt the canonical motion to the physical structure and dynamics of the target character. This process mitigates physical artifacts and naturally stylizes the motion to reflect the unique morphology of each character.

Our main contributions are as follows:
\begin{itemize}
\item We present a character-agnostic in-betweening framework that removes the dependency on a fixed skeleton and enables generalization across characters with diverse body types and proportions.

\item We introduce a physics-based motion adaptation stage for motion in-betweening that eliminates artifacts and stylizes the motion to reflect each character's unique morphology and physical characteristics.
\end{itemize}

We validate our approach on both standard benchmarks and stylized characters. Experiments show that our method produces physically plausible, stylistically consistent, and keyframe-respecting motions across a wide range of constraints.

\section{Related Work}

\subsection{Physically-based Character Animation}

Physically-based character animation aims to synthesize motions that are both visually realistic and physically plausible by simulating underlying dynamics such as contact, mass, and balance. Compared to kinematic approaches, physics-based methods offer improved realism, particularly in scenarios involving sparse observations or interactive environments.

Early methods relied on handcrafted controllers and trajectory optimization \cite{witkin1988spacetime,liu2005learning}, but lacked scalability. With the rise of deep reinforcement learning (RL), physics-based control became more generalizable. DeepMimic \cite{peng2018deepmimic} demonstrated that RL policies can robustly imitate complex motion clips, inspiring further work on skill learning and motion adaptation.
QuestSim \cite{winkler2022questsim} extended this idea to sparse-input tracking, utilizing only headset and controller data to control a full-body avatar in a physics simulator. Reda et al.\cite{reda2023physics} further generalized this to real-time motion retargeting for characters with varying morphologies using the same sparse input, enabling cross-skeleton adaptation.
MaskedMimic \cite{tessler2024maskedmimic} proposed a unified control framework that learns to inpaint motion segments from diverse control signals (e.g., text, keyframes) in a physics simulation. This allows seamless task-switching under a single model while maintaining physical realism.

Beyond RL, Fussell et al.\cite{fussell2021supertrack} introduced SuperTrack, a supervised learning approach that replaces PPO with a differentiable physics model to directly optimize tracking policies. This improves training stability and reduces the sample inefficiency typically associated with model-free RL.

Recently, physics-based priors have also been integrated into generative models. PhysDiff \cite{yuan2023physdiff} introduced a diffusion framework that incorporates physics projection during denoising, producing motions with fewer artifacts such as foot sliding and ground penetration. Unlike post-processing approaches, this integration ensures physical consistency throughout the generative process.

Together, these works highlight the importance of physics-based modeling in synthesizing realistic human motion. 
Despite significant progress, most approaches are restricted to a fixed skeletal structure
and few offer generative capabilities for novel motions from sparse keyframes. This motivates the integration of physics-based adaptation with generative motion models, as explored in this work.

\subsection{Diffusion Models}

Diffusion models have emerged as a powerful class of generative models, achieving state-of-the-art performance across a wide range of domains, including image synthesis~\cite{ramesh2022hierarchical,dhariwal2021diffusion,rombach2022high,saharia2022photorealistic}, video generation~\cite{ho2022imagen,ho2022video}, 3D shape modeling~\cite{luo2021diffusion,zhou20213d}, and speech synthesis~\cite{kong2020diffwave}. Compared to traditional generative models such as GANs and VAEs, diffusion models offer notable advantages, including improved sample diversity, enhanced training stability, and better generalization. These properties make them particularly well-suited for modeling complex, high-dimensional temporal data.

A core feature of diffusion models is their iterative generation process, which refines samples from Gaussian noise through a discrete Markov chain or a continuous stochastic differential equation (SDE) formulation~\cite{song2020score}. The Denoising Diffusion Probabilistic Model (DDPM)~\cite{ho2020denoising} established the foundation of this approach, demonstrating that high-quality samples can be generated by reversing a noise-injection process. Subsequent works have introduced various improvements to enhance efficiency and controllability. Classifier-free guidance~\cite{ho2022classifier} improves conditional generation, while latent diffusion models (LDMs)~\cite{rombach2022high} reduce computational overhead by operating in compressed latent spaces. Other advances, such as flow matching~\cite{lipman2022flow}, score distillation sampling (SDS)~\cite{poole2022dreamfusion}, and denoising diffusion implicit models (DDIMs)~\cite{song2020denoising}, further enhance generation quality and inference speed.

In the field of human motion synthesis, diffusion models have been widely adopted for tasks such as motion generation~\cite{dabral2023mofusion,ma2022pretrained,zhou2023ude,kim2023flame} and character control~\cite{chen2024taming}. Notable frameworks like Motion Diffusion Model (MDM)~\cite{tevet2022human} and MotionDiffuse~\cite{zhang2024motiondiffuse} leverage diffusion to model complex motion distributions, capturing diverse motion styles and improving temporal coherence. Recent extensions incorporate language encoders for text-conditioned motion synthesis~\cite{tevet2022human,chen2023executing,shafir2024human,zhang2024motiondiffuse}, enabling semantically meaningful and controllable animation. 
These methods highlight the potential of diffusion models in generating context-aware, realistic human motion. However, most existing works focus on purely kinematic generation and do not account for physical consistency, limiting their applicability when motions are transferred to characters with distinct physical properties.

\subsection{Motion In-betweening}

Motion in-betweening refers to the task of generating transitions between fixed past and future keyframes. It is often formulated as a constrained motion prediction problem.
A common approach involves interpolating keyframes using spline-based techniques such as Bézier curves, B-splines, or Catmull-Rom splines. While effective for producing smooth trajectories, these methods are labor-intensive to tune and often lack physical realism.
To ease this process, Ciccone et al.~\cite{ciccone2019tangent} proposed automatically optimizing spline tangents using an intuitive trajectory-based control interface, reducing the need for manual tangent tuning.

Physically-based methods improve realism by formulating motion generation as a spacetime optimization problem~\cite{witkin1988spacetime, ngo1993spacetime}, where trajectories are computed to satisfy both keyframe constraints and physical laws. Witkin and Popović~\cite{witkin1995motion} introduced a framework for editing existing motions through curve warping, enabling plausible modifications while preserving physical validity. Gleicher~\cite{gleicher1997motion} extended this concept by incorporating spacetime constraints directly into the editing process, allowing for more physically consistent motion retargeting. 
In parallel, statistical models offer a data-driven alternative to hand-tuned interpolation. MAP-based techniques~\cite{chai2007constraint, min2009interactive}, Gaussian process models~\cite{wang2007gaussian}, and Markov models~\cite{lehrmann2014efficient} have been applied to learn motion distributions and synthesize transitions. These models better capture stylistic variations than spline interpolation, but often struggle with scalability when applied to large datasets.

With the rise of deep learning, recent work has shifted toward deep learning-based approaches, which have demonstrated stronger performance in modeling complex, diverse, and context-aware transitions.
Harvey and Pal \cite{harvey2018recurrent} introduced Recurrent Transition Networks (RTN) using LSTM-based autoregressive models, later improved with adversarial training and time-to-arrival embeddings for better keyframe alignment \cite{harvey2020robust}. 
CNN-based methods \cite{hernandez2019human,kaufmann2020convolutional} leveraged fully convolutional autoencoders to complete missing frames, though they faced challenges in modeling long-term dependencies due to limited receptive fields.
Transformer-based architectures have demonstrated better performance due to their ability to capture global temporal relationships \cite{duan2021single,qin2022motion}. Duan et al. \cite{duan2021single} proposed a non-autoregressive in-betweening framework using a BERT-style Transformer, while Qin et al. \cite{qin2022motion} introduced a two-stage Transformer framework, where a Context Transformer generates coarse transitions, refined by a Detail Transformer. Despite these advances, non-generative models often struggle to produce diverse motion outputs when constrained by the same keyframes.

To address this limitation, recent approaches have incorporated diffusion models into motion in-betweening to improve both diversity and realism. PhysDiff~\cite{yuan2023physdiff} introduces physical constraints during the denoising process to enhance motion plausibility, but lacks explicit keyframe control. 
Guided Motion Diffusion (GMD)~\cite{karunratanakul2023guided} improves keyframe adherence by integrating trajectory and keyframe constraints, enabling better transitions with sparse inputs.
However, many diffusion-based methods still impose keyframe constraints only at inference time, which can lead to inconsistencies with the keyframes, especially when keyframes are sparse. 
To overcome this, DiffKFC~\cite{wei2024enhanced} incorporates keyframe conditioning into training, improving keyframe adherence while maintaining generative diversity.  
Qin et al.~\cite{qin2024robust} further introduce a lightweight Transformer-based framework that also integrates keyframe supervision into training. 
Their method achieves better adherence to keyframe constraints under various keyframe densities and introduces evaluation metrics such as K-FID, K-Diversity, and K-Error to systematically assess the quality of generative in-betweening methods.

Building on these insights, our method extends diffusion-based motion in-betweening with a two-stage framework. A character-agnostic diffusion model first generates motion on a canonical skeleton, enabling reuse across diverse characters without retraining. 
This is followed by a physics-based adaptation stage that transfers the motion to the target character, ensuring physical plausibility and style consistency.
Our framework supports scalable, physically grounded, and style-consistent motion in-betweening generation under sparse keyframe constraints.

\section{Background}
\subsection{Diffusion Models}
Diffusion models are a class of generative models that learn to approximate complex data distributions by gradually denoising samples from a known noise distribution. The modeling process consists of two stages: a forward diffusion process that adds noise to data, and a reverse process that reconstructs data from noise.
Let $\mathbf{x}_0 \sim q(\mathbf{x})$ be a data point sampled from the real data distribution. The forward process defines a Markov chain that progressively adds Gaussian noise over $T$ steps to obtain a sequence of latent variables $\mathbf{x}_1, \dots, \mathbf{x}_T$, where each step is defined as:
\begin{equation}
q(\mathbf{x}_t \mid \mathbf{x}_{t-1}) = \mathcal{N}\left( \mathbf{x}_t; \sqrt{1 - \beta_t} \, \mathbf{x}_{t-1}, \beta_t \mathbf{I} \right)
\end{equation}
 with a predefined noise schedule $\left\{ \beta_t \right\}_{t=1}^T
$, where $\beta_t \in (0, 1)$. The full forward process is expressed as: 
\begin{equation}
q(\mathbf{x}_{1:T} \mid \mathbf{x}_0) = \prod_{t=1}^T q(\mathbf{x}_t \mid \mathbf{x}_{t-1})
\end{equation}

As $t$ increases, the sample $\mathbf{x}_t$ becomes increasingly noisy. When $t \to T$, $\mathbf{x}_t$ converges to an isotropic Gaussian distribution: 
\begin{equation} 
\mathbf{x}_T \sim \mathcal{N}(\mathbf{0}, \mathbf{I}). \end{equation}

To generate new data, the model learns to reverse this process that gradually denoises a Gaussian sample back to the data manifold. Assuming $q(\mathbf{x}_{t-1} \mid \mathbf{x}_t)$ is also Gaussian, we approximate it using a learnable model $p_\theta$:
\begin{equation}
p_\theta(\mathbf{x}_{0:T}) = p(\mathbf{x}_T) \prod_{t=1}^T p_\theta(\mathbf{x}_{t-1} \mid \mathbf{x}_t)
\end{equation}
\begin{equation}
p_\theta(\mathbf{x}_{t-1} \mid \mathbf{x}_t) = \mathcal{N}\left( \mathbf{x}_{t-1}; \mu_\theta(\mathbf{x}_t, t), \Sigma_\theta(\mathbf{x}_t, t) \right)
\end{equation}

where $\mu_\theta(\mathbf{x}_t, t)$ and $\Sigma_\theta(\mathbf{x}_t, t)$ are parameterized using neural networks or derived from network predictions .

This generative formulation enables the model to recover complex data samples from pure noise via iterative denoising. In practice, the network is trained to predict either the noise $\epsilon$ added during the forward process or the original data $\mathbf{x}_0$ \cite{song2020score,ho2020denoising}.

\subsection{Reinforcement Learning}

We adopt the reinforcement learning (RL) framework to formalize sequential decision-making over motion states. In this setting, the agent interacts with the environment over discrete time steps to maximize long-term rewards. The process is modeled as a Markov Decision Process (MDP) defined by the tuple $(\mathcal{S}, \mathcal{A}, P, R, \gamma)$, where $\mathcal{S}$ denotes the state space, $\mathcal{A}$ the action space, $P(s_{t+1} \mid s_t, a_t)$ the transition probability, $R(s_t, a_t)$ the reward function, and $\gamma \in [0, 1]$ the discount factor.

The agent follows a policy $\pi_\theta(a_t \mid s_t)$ parameterized by $\theta$, which defines the probability of taking action $a_t$ in state $s_t$. Given a trajectory $\tau = (s_0, a_0, r_1, \dots, s_T)$ sampled from the policy, the expected discounted return is: \begin{equation} J(\theta) = \mathbb{E}_{\tau \sim P_\theta(\tau)} \left[ \sum_{t=0}^T \gamma^t r_t \right]. \end{equation} Here, each reward $r_t$ corresponds to the immediate reward obtained by taking action $a_t$ in state $s_t$, typically defined as $r_t = R(s_t, a_t)$.
The trajectory distribution under policy $\pi_\theta$ is defined as: \begin{equation} P_\theta(\tau) = P(s_0) \prod_{t=0}^{T} P(s_{t+1} \mid s_t, a_t)\pi_\theta(a_t \mid s_t), \end{equation} where $P(s_0)$ is the initial state distribution.
Together, the MDP structure and return objective define a principled foundation for optimizing sequential decision-making in our framework.

\section{Methodology}
\subsection{Overview}
\begin{figure}[thb] 
    \centering
    \includegraphics[width=0.9\textwidth]{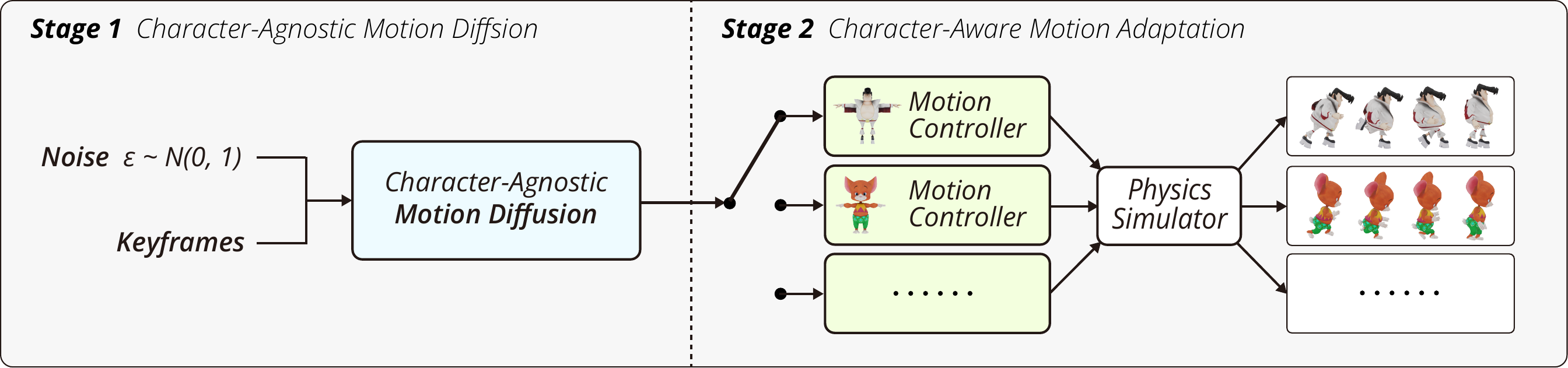}
    \caption{Overview of the framework.}
    \label{fig:overview}
\end{figure}

This work presents a robust framework for generating motion in-betweening. It is capable of producing animations for different characters using a standardized character motion dataset, without requiring various datasets for different body types or skeletal structures. Our approach, as depicted in Figure \ref{fig:overview}, follows a two-stage process.

In the first stage, \emph{Character-agnostic Motion Diffsion}, we develop a motion diffusion model leveraging the Transformer architecture, trained on an extensive motion capture dataset based on a canonical skeleton. 
This model takes keyframes and a noise vector as inputs and progressively refines the motion through a denoising process. 
Its goal is to generate realistic in-betweening motions while maintaining consistency with the given keyframes. It is capable of handling a wide range of motion transitions while allowing for variations in the generated results.

In the second stage, \emph{Character-Aware Motion Adaptation}, we train a motion controller using reinforcement learning to adapt the generated motions to specific character models. The controller learns a policy that outputs proportional-derivative (PD) controller targets for a physics simulator, ensuring that the motion adheres to physical constraints. By maintaining physical plausibility, the controller naturally captures the unique style of each character, as their morphology and dynamics influence the resulting motion. 

This approach leverages the diverse and flexible motion generation capabilities of the kinematic model from Stage 1 while incorporating the physical realism of Stage 2 to adapt motions to specific characters. By integrating these two stages, our framework eliminates the need for manually curated motion data for each character, enabling generalization across different skeleton structures and body types while maintaining both control flexibility and physical plausibility.

\subsection{Character-agnostic Motion Diffusion}
\subsubsection{Network Architecture}

We follow the motion diffusion architecture introduced in \cite{qin2024robust} for the Character-Agnostic Motion Diffusion stage. The network features Keyframe-Masked Transformer layers that effectively enforce keyframe constraints during motion generation.

We represent a motion clip as $\mathbf{x} = \{x_1, x_2, \ldots, x_N \}$, where $N$ is the window length.  
Each frame $x_n$ contains pose information for a skeleton with $J$ joints: 
$x_n = \{ r_n, v_n^x, p_n^y, v_n^z, u_n \}$. 
Here, $r_n \in \mathbb{R}^{J \times 6}$ denotes the 6D local rotations of all joints~\cite{zhou2019continuity}. 
The root joint's motion is represented by its velocity in the $x$ and $z$ directions ($v_n^x$, $v_n^z$), and its absolute position along the $y$-axis ($p_n^y$). An additional velocity term $u_n \in \mathbb{R}^{6 + 4 \times 3}$ captures the root’s angular velocity as well as the global velocities of the hands and feet on both sides. 
The sparse keyframes are denoted as $\mathbf{x^{\text{kf}}} = \{x_{k_1}, x_{k_2}, \ldots, x_{k_K} \}$, where $\{k_i\}$ indicates their positions within the full motion clip $\mathbf{x}$.  
The remaining frames are treated as in-betweening frames, defined as $\mathbf{x^{\text{ib}}} = \mathbf{x} \setminus \mathbf{x^{\text{kf}}}$. 
A binary mask $\mathbf{M}$ is used to indicate keyframes, where zeros correspond to the specified keyframe indices.
All features are normalized using the mean and standard deviation computed from the training set. 
Finally, the input to the network at diffusion timestep $t$ is defined as  
\begin{equation}
\tilde{\mathbf{x}}_t = \mathbf{x}_t \odot \mathbf{M} + \mathbf{x}_0 \odot (1 - \mathbf{M}),
\end{equation}
where the noisy motion clip $\mathbf{x}_{t} \sim q(\mathbf{x}_{t} \mid \mathbf{x}_0)$, $\mathbf{x}_0$ is the clean (noise-free) motion clip, $\odot$ denotes the Hadamard (element-wise) product, and $\tilde{\mathbf{x}}_t \in \mathbb{R}^{N \times D}$ with $D = J \times 6 + 21$.

The network adopts a lightweight encoder–transformer–decoder architecture composed of an MLP encoder, $L$ Keyframe-Masked Transformer layers, and an MLP decoder. This design enables the enforcement of keyframe constraints while supporting long-range in-betweening generation.
The input to the network at timestep $t$ is first encoded using an MLP:

\begin{equation}
\mathbf{\tilde{h}} = \operatorname{MLP}(\tilde{\mathbf{x}}_{t})
\end{equation}

The timestep $t$ is embedded via a sinusoidal encoding followed by another MLP, and concatenated with the motion embedding:

\begin{equation}
\mathbf{h}^{0} = \operatorname{Concatenate}(\mathbf{\tilde{h}}, \operatorname{MLP}(\operatorname{Sinusoidal}(t)))
\end{equation}

To ensure that keyframes are preserved during the reverse diffusion process, a binary mask $\mathbf{M}$ is applied to modulate each Transformer layer. Following~\cite{qin2024robust}, we multiply the outputs of both self-attention and feed-forward submodules with $\mathbf{M}$, effectively disabling updates to keyframe positions and allowing them to pass through unchanged. This design encourages identity mapping behavior for clean keyframes while allowing denoising of noisy frames.
The latent features are refined through $L$ Keyframe-Masked Transformer layers as follows:

\begin{equation}
\label{eqn:mhsa}
\mathbf{\tilde{h}}^{l} = \mathbf{h}^{l-1} + \operatorname{MHSA}(\operatorname{LayerNorm}(\mathbf{h}^{l-1}), \mathbf{pe}_{rel}) \odot \mathbf{M}
\end{equation}
\begin{equation}
\label{eqn:pffn}
\mathbf{h}^{l} = \mathbf{\tilde{h}}^{l} + \operatorname{PFFN}(\operatorname{LayerNorm}(\mathbf{\tilde{h}}^{l}) \odot \mathbf{M}
\end{equation}

where $l = 1, 2, \ldots, L$.
To generalize beyond the training clip length, we incorporate the learned relative positional encoding $\mathbf{pe}_{rel}$~\cite{qin2022motion}, which encodes frame distances invariant to absolute positions. We implement it using the skew mechanism proposed by~\cite{huang2018music}.
Finally, the denoised motion at timestep $t$ is predicted by decoding the last layer’s output:

\begin{equation}
\hat{\mathbf{x}}_{t} = \operatorname{MLP}(\mathbf{h}^L)
\end{equation}

\begin{figure}[t] 
    \centering
    \includegraphics[width=\textwidth]{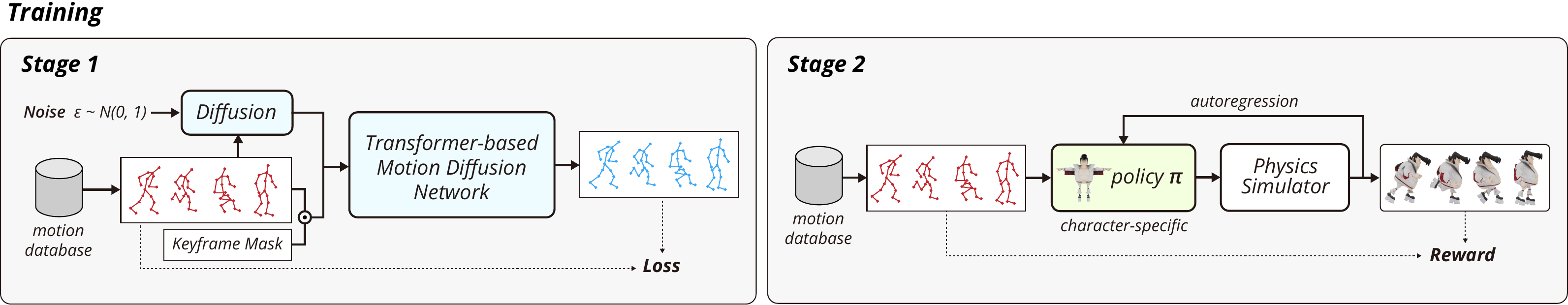}
    \caption{The training pipeline.}
    \label{fig:train_pipe}
\end{figure}

\subsubsection{Training and Inference}

The overall training pipeline is illustrated in Figure~\ref{fig:train_pipe}. At each training step, keyframe indices are randomly sampled from the input sequence. The number of keyframes is controlled by a hyperparameter termed the \textit{constrained rate}, which defines the maximum allowed ratio of keyframes to the full clip length. In some cases, no keyframes are provided, encouraging the model to generalize to unconstrained scenarios.

Since the model predicts clean motion directly instead of noise, we incorporate geometric losses to constrain poses, prevent artifacts, and encourage natural and coherent motions. The training objective includes an $\ell_2$ reconstruction loss:

\begin{equation}
    \mathcal{L}_{simple} = E_{\mathbf{x}_0\sim q\left(\mathbf{x}_0 \mid \mathbf{x}_0^{kf}\right),  t \sim[1, T]}\left[\left\|\mathbf{x}_0-\mathbf{x}_\theta\left(\mathbf{x}_t^{ib}, \mathbf{x}_0^{kf}, t\right)\right\|_2^2\right]
\end{equation}

In addition, we apply an $\ell_1$ loss on local joint rotations:

\begin{equation}
\mathcal{L}_{rot} = \frac{1}{N}\sum_{n=1}^N\left\lVert r_n - \hat{r}_n\right\rVert_1
\end{equation}

To reflect the hierarchical nature of the skeleton, we further apply an $\ell_1$ loss to global joint positions, derived via forward kinematics:

\begin{equation}
\mathcal{L}_{pos} = \frac{1}{N}\sum_{n=1}^N\left\lVert g_n - \hat{g}_n \right\rVert_1,\: g_n = FK(r_n, v_n^x, p_n^y, v_n^z),
\end{equation}

where $g_n \in \mathbb{R}^{J \times 3}$ denotes the global joint positions at frame $n$. The function $FK(\cdot)$ computes global positions from local rotations and root motion. The final objective is a weighted sum of all loss terms:

\begin{equation}
    \mathcal{L}=\lambda_{simple} \mathcal{L}_{simple}+\lambda_{rot} \mathcal{L}_{rot}+\lambda_{pos} \mathcal{L}_{pos},
\end{equation}
where $\lambda_{\text{simple}} = 1.0$, $\lambda_{\text{rot}} = 1.0$, and $\lambda_{\text{pos}} = 0.02$.

\begin{figure}[tb] 
    \centering
    \includegraphics[width=\textwidth]{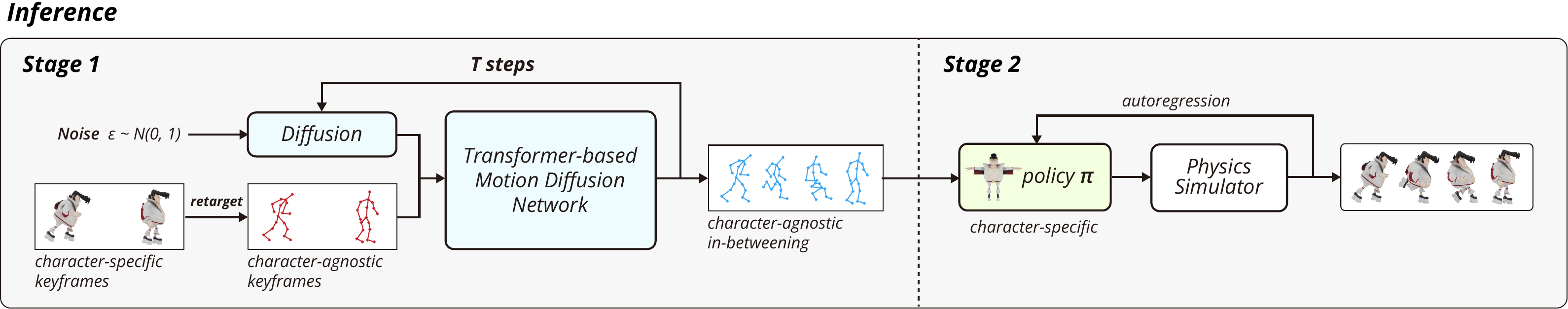}
    \caption{The inference pipeline.}
    \label{fig:infer_pipe}
\end{figure}

At inference time, as illustrated in Figure~\ref{fig:infer_pipe}, the model takes a set of character-specific keyframes as input.  
Since the diffusion model is trained on a canonical skeleton, these keyframes are first retargeted to the canonical form to ensure compatibility.  
This is achieved using a predefined joint mapping that transfers global joint rotations from the original character to the canonical skeleton.  
For root joint positions, we apply a scaling factor based on the ratio of hip heights between the source and canonical characters, resulting in character-agnostic keyframes suitable for the diffusion model.

Given the retargeted keyframes, the model then synthesizes the in-between motion by performing denoising starting from a Gaussian noise vector $\mathbf{x}_T^{ib} \sim \mathcal{N}(0, \mathbf{I})$.  
Conditioned on the keyframe constraints $\mathbf{x}_0^{kf}$, the diffusion model iteratively refines the noisy input using a learned reverse process:

\begin{equation}
p_\theta\left(\mathbf{x}_{t-1}^{ib} \mid \mathbf{x}_t^{ib}, \mathbf{x}_0^{kf}\right) = \mathcal{N}\left(\mathbf{x}_{t-1}^{ib}; \mu_\theta\left(\mathbf{x}_t^{ib}, \mathbf{x}_0^{kf}, t\right), \sigma_t^2 \mathbf{I} \right),
\end{equation}

where $\mu_\theta$ is predicted by the Transformer-based denoising network, and $\sigma_t^2$ is a pre-defined variance schedule.  
After $T$ denoising steps, the final character-agnostic motion $\mathbf{x}_0^{ib}$ is produced, which is then passed to Stage~2 for character-specific adaptation.

\subsection{Character-Aware Motion Adaptation}
\begin{figure}[htb] 
    \centering
    \includegraphics[width=0.8\textwidth]{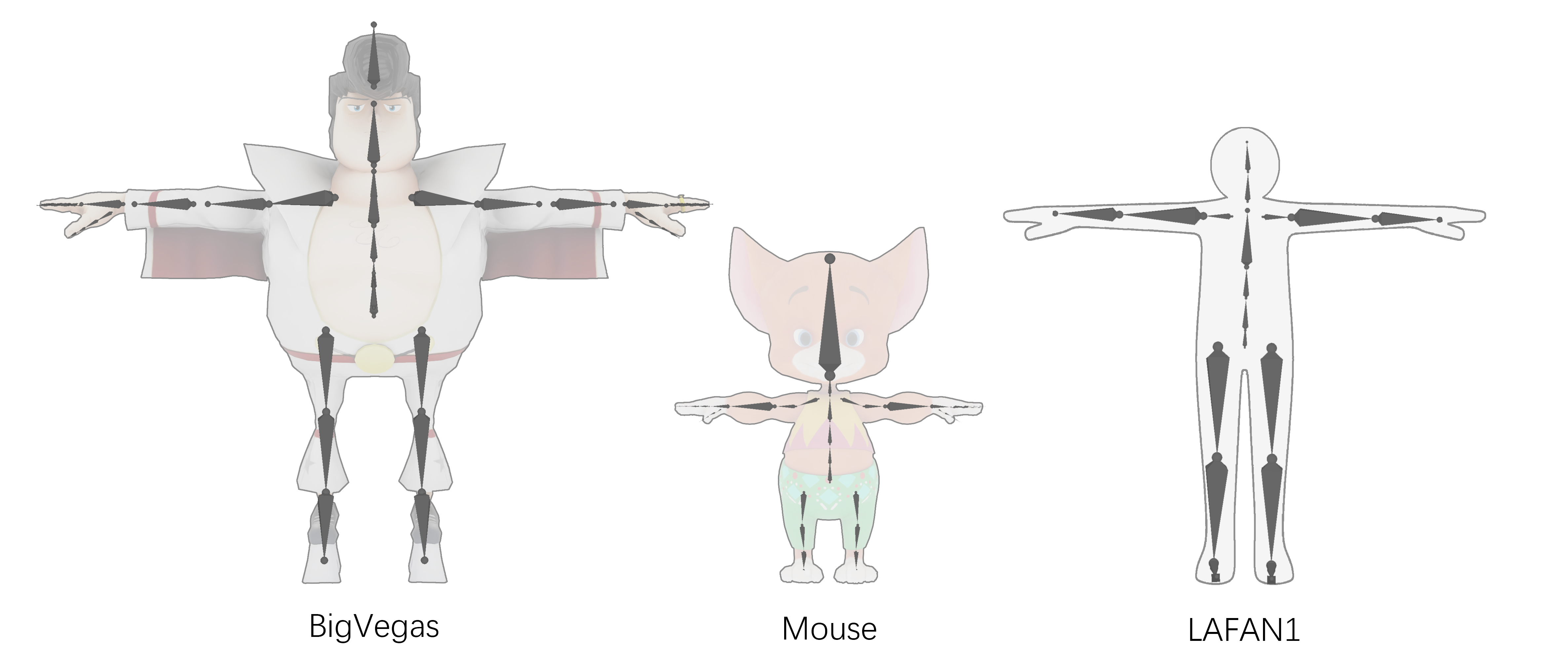}
    \caption{Examples of characters with different body proportions and skeletal structures.}
    \label{fig:char_skel}
\end{figure}

\begin{figure}[htb] 
    \centering
    \includegraphics[width=0.4\textwidth]{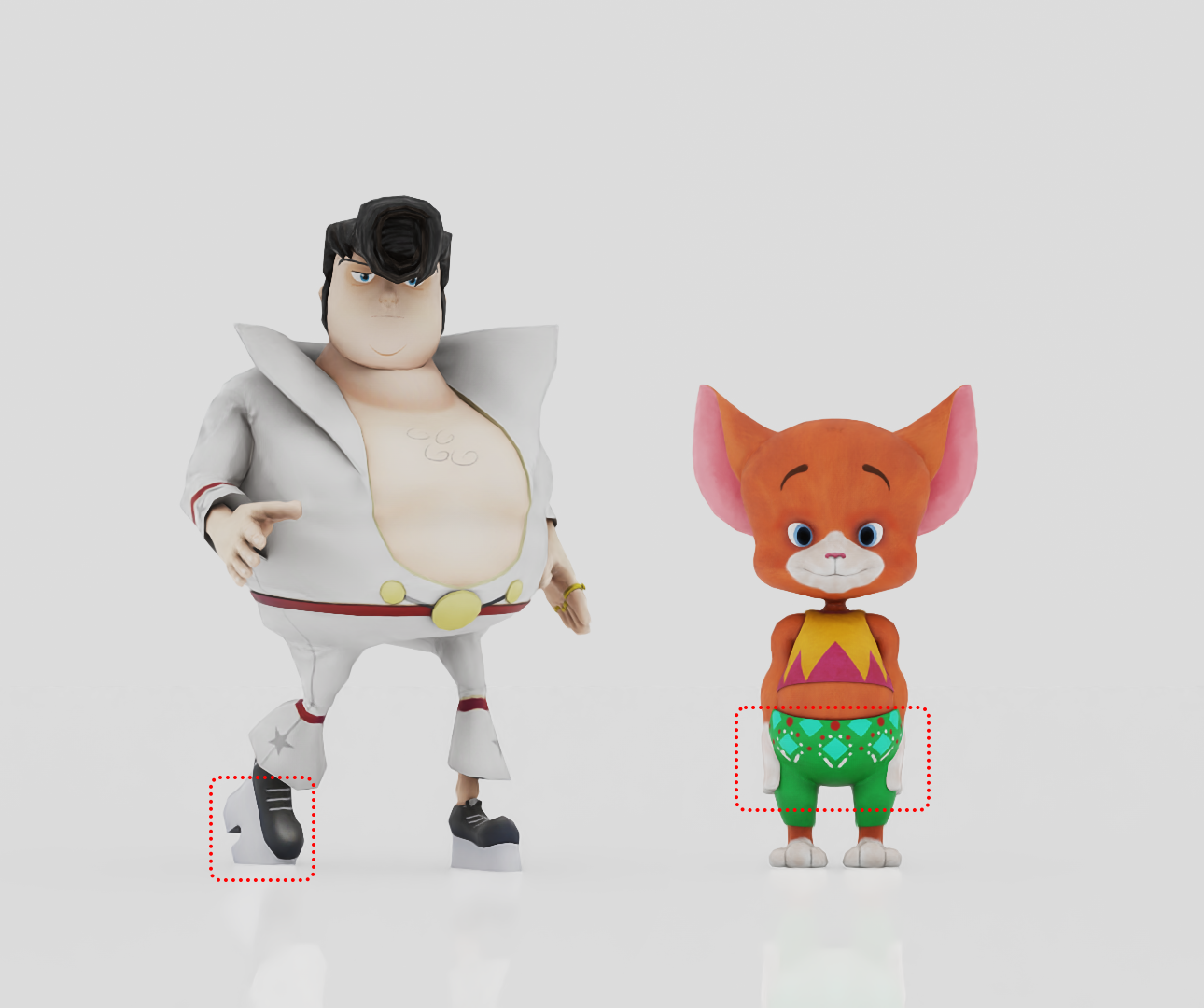}
    \caption{Visual artifacts resulting from direct kinematic retargeting. The discrepancies between character body shapes and the motion data used for training cause interpenetration artifacts, as indicated by the red boxes.}
    \label{fig:artifacts}
\end{figure}

The previous stage is trained and inferred on character-agnostic motion data.
However, in practice, characters may vary considerably in both body proportions and skeletal structures.
As shown in Figure~\ref{fig:char_skel}, the stylized characters (BigVegas and Mouse~\cite{mixamo}) have skeletal structures that differ greatly from that of the training dataset LAFAN1~\cite{harvey2020robust}. 
Directly applying the motion generated from the previous stage to these characters via kinematic retargeting can lead to noticeable artifacts, such as foot sliding or mesh interpenetration (see Figure~\ref{fig:artifacts}).

To address the limitations of kinematic retargeting, we adopt a physics-based motion adaptation strategy in Stage 2. For each character, we train a motion controller using reinforcement learning to reproduce the target motion generated in Stage 1. This policy $\pi$ learns to produce physically plausible movements that reflect the character’s unique morphology and dynamics. 
Given the wide variation in body proportions and skeletal structures across characters, our proposed method provides a solution for adapting motion in a physically consistent and style-preserving manner.

In the following, we detail the reinforcement learning setup, including the observation space, action space, reward design, and training procedure used to build these character-specific controllers.

\subsubsection{Observation Space}
\begin{figure}[htb] 
    \centering
    \includegraphics[width=\textwidth]{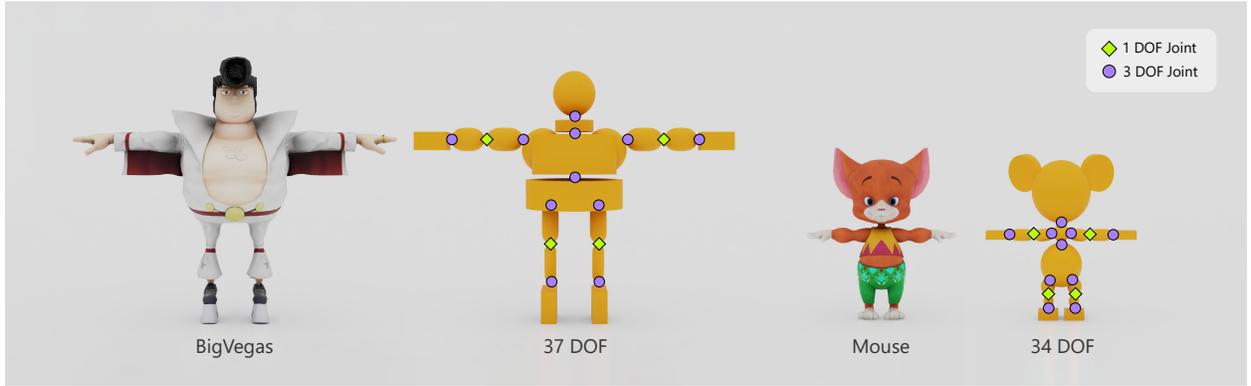}
    \caption{Physically simulated characters with different numbers of actuated degrees of freedom (DOFs).}
    \label{fig:char_rb}
\end{figure}

At each frame \( n \), the motion control policy \( \pi \) receives an observation \( \mathbf{o}_n = \{s_n, t_n\} \), which encodes both the current physical state of the simulated character (illustrated in Figure~\ref{fig:char_rb}) and the future motion to be tracked.

The character's state \( s_n \) includes its pose, position, and velocity, all canonicalized with respect to the local coordinate frame of the root joint. Specifically, the observation consists of joint positions \( p_n \), local joint rotations \( r_n \), joint velocities \( \dot{p}_n \), and angular velocities \( \dot{r}_n \), each expressed relative to the root:
\begin{equation}
s_n = \left\{
(p_n - p_n^{\text{root}}) \ominus r_n^{\text{root}},\ 
r_n \ominus r_n^{\text{root}},\ 
\dot{p}_n \ominus r_n^{\text{root}},\ 
\dot{r}_n \ominus r_n^{\text{root}}
\right\},
\end{equation}
where \( \ominus \) denotes a relative transformation with respect to the root's orientation, ensuring that the observation remains invariant to global translation and heading.

In addition to the current state, the observation also includes the next target pose \( t_n \) from the reference motion. It is encoded relative to the same canonicalized coordinate frame of the current pose:
\begin{equation}
t_n = \left\{
r_{n+1} \ominus r_n^{\text{root}},\ 
(p_{n+1} - p_n) \ominus r_n^{\text{root}}\right\},
\end{equation}
where \( r_{n+1} \) and \( p_{n+1} \) denote the joint rotations and positions of the next target pose.

\subsubsection{Action Space}
The action space consists of target joint positions predicted by the policy \( \pi \) at each timestep. These targets are executed by a proportional-derivative (PD) controller, following the setup in~\cite{peng2018deepmimic}. The dimensionality of the action space depends on the number of actuated degrees of freedom (DOFs) of the character. As illustrated in Figure~\ref{fig:char_rb}, BigVegas and Mouse have 37 and 34 DOFs respectively. Each action \( a_n \) is sampled from a multivariate Gaussian distribution with a fixed diagonal covariance, where the standard deviation is set to \( \sigma^\pi = 0.05 \). During inference, we take the mean of this distribution as the final action output.

\subsubsection{Rewards}
Our reward function encourages the simulated character to track the reference motion produced by the diffusion model. 
Although the simulated and reference characters have different skeletons and joint configurations, 
we leverage the same predefined joint mapping used during keyframe retargeting to establish correspondence between them. 
The reward is then computed over the mapped joints, enabling the simulated character to follow the canonical motion trajectory in a physically consistent manner, while allowing broad compatibility across characters with varying morphologies.

Specifically, we include imitation terms that penalize differences in global joint positions (\(gp\)), global joint rotations (\(gr\)), global linear velocities (\(g\dot{p}\)), global angular velocities (\(g\dot{r}\)), and apply an energy penalty (\(eg\)) to encourage smoother motions~\cite{lee2023questenvsim}.
The overall reward at frame \( n \) is defined as a weighted sum of the following components, each formulated using a Gaussian kernel:
\begin{equation}
r_n = 
w_{gp} \, r_n^{gp} + 
w_{gr} \, r_n^{gr} + 
w_{g\dot{p}} \, r_n^{g\dot{p}} + 
w_{g\dot{r}} \, r_n^{g\dot{r}} + 
w_{eg} \, r_n^{eg},
\end{equation}
\begin{equation}
r_n^{(\cdot)} = \exp\left( -k^{(\cdot)} \, \| f_n^{(\cdot)} - \hat{f}_n^{(\cdot)} \|^2 \right),
\end{equation}
where \( f_n^{(\cdot)} \) and \( \hat{f}_n^{(\cdot)} \) are the corresponding feature vectors from the simulated character and the mapped reference pose. The weights are set to \( w_{gp} = 0.8 \), \( w_{gr} = 1.0 \), \( w_{g\dot{p}} = 0.2 \), \( w_{g\dot{r}} = 0.2 \), and \( w_{eg} = 0.001 \), and the sensitivity parameters are \( k_{gp} = 10 \), \( k_{gr} = 50 \), \( k_{g\dot{p}} = 1 \), \( k_{g\dot{r}} = 1 \), and \( k_{eg} = 0.01 \).

As mentioned above, due to differences in skeletal structure, directly comparing global joint positions or velocities between the simulated character and the reference motion is not meaningful. 
Instead, we leverage the same joint mapping used during keyframe retargeting to transfer motion information from the reference skeleton to the simulated character. Specifically, we copy global joint rotations from the reference motion to the corresponding joints of the simulated character according to the predefined mapping. The root position is also adjusted by scaling it according to the ratio of hip heights between the reference and simulated skeletons.
We then extract the global joint positions and velocities from the resulting pose on the simulated character, and use these as reference features for computing the reward.
This ensures consistent evaluation across characters with different morphologies and skeletal configurations, allowing the policy to learn motion tracking behavior that generalizes across diverse character embodiments.

\subsubsection{Training and Inference}
Similar to the previous stage, our motion controller policy is trained on subjects 1–4 of the LAFAN1 dataset~\cite{harvey2020robust}. Since LAFAN1 includes motions such as stair climbing and obstacle crossing, we manually annotate the positions of relevant environmental colliders. Without these colliders, the reference motions become physically invalid, making it impossible for the simulated character to track them accurately. 

To ensure that the simulated character interacts with the environment in a manner consistent with the reference motion, the colliders are uniformly scaled in world coordinates during training, using the same hip-height ratio as in the root position retargeting strategy described earlier.

We train a policy \( \pi \) for each character through reinforcement learning. As shown in Figure~\ref{fig:train_pipe} (Stage 2), a reference motion clip is sampled at the beginning of each episode, and its frames are sequentially fed into the policy. The policy outputs actuation targets to drive the simulated character, generating motion auto-regressively until the episode ends or early termination is triggered~\cite{peng2018deepmimic}. The reward is computed based on the discrepancy between the simulated and reference motion.

We implement our networks using PyTorch~\cite{paszke2019pytorch} and train them with the Proximal Policy Optimization (PPO) algorithm~\cite{schulman2017proximal}. Both the policy and value functions are modeled as multilayer perceptrons with three hidden layers of sizes [1024, 512, 256] and ReLU activations. We set the PPO clip ratio to 0.2. Advantages are estimated using Generalized Advantage Estimation (GAE) with \(\lambda = 0.95\) and a discount factor \(\gamma = 0.99\). The networks are optimized using the Adam optimizer~\cite{kingma2014adam} with \(\beta_1 = 0.9\), \(\beta_2 = 0.999\), and a learning rate of 1e\text{-}4.

All simulations are conducted in IsaacLab~\cite{mittal2023orbit} at 30 FPS. We simulate 4096 characters in parallel, each performing 15 steps under the same policy, resulting in 61,440 transition samples per iteration. The models are trained for a total of 10 billion policy steps on a single NVIDIA RTX 3080 Ti GPU.

At inference time (Figure~\ref{fig:infer_pipe}, Stage 2), the policy takes as input the motion generated by the previous stage. The final motion is produced in an auto-regressive manner through recurrent interaction with the physics simulator.

\section{Experiments}
\subsection{Datasets and Metrics}
We conduct our experiments on the LAFAN1 dataset~\cite{harvey2020robust}, which contains 496,672 frames of motion capture data recorded at 30 FPS, covering a range of human actions performed by five subjects. Following standard practice, sequences from Subjects 1–4 are used for training, while Subject 5 is held out for evaluation.

To assess in-betweening performance under sparse keyframe constraints, we adopt three evaluation metrics proposed by Qin et al.~\cite{qin2024robust}, which measure motion quality, diversity, and keyframe consistency.

\textbf{K-FID} measures the distribution discrepancy between the generated motion and the ground truth based on per-frame latent representations extracted using a pretrained Transformer-based autoencoder:
\begin{equation}
\text{K-FID}  = \left\lVert \mu - \hat{\mu} \right\rVert_2^2 + tr(\Sigma + \hat{\Sigma} - 2(\Sigma \hat{\Sigma})^{\frac{1}{2}})
\end{equation}

\textbf{K-Diversity} quantifies the variation within each generated motion sequence, where $(i, j)$ are N pairs of random frame indices:
N indices pairs :
\begin{equation}
\text{K-Diversity} = \frac{1}{N}\sum_{i=1}^N {\left\lVert z_i - z_j\right\rVert_2}
\end{equation}

\textbf{K-Error} computes the mean global position error at keyframe joints between the prediction and ground truth:
\begin{equation}
\text{K-Error} = \frac{1}{J}\frac{1}{K}\sum_{i=1}^K {\left\lVert g_{k_i} - \hat{g}_{k_i} \right\rVert_2}
\end{equation}

\begin{figure}[th]
    \centering
    \includegraphics[width=\textwidth]{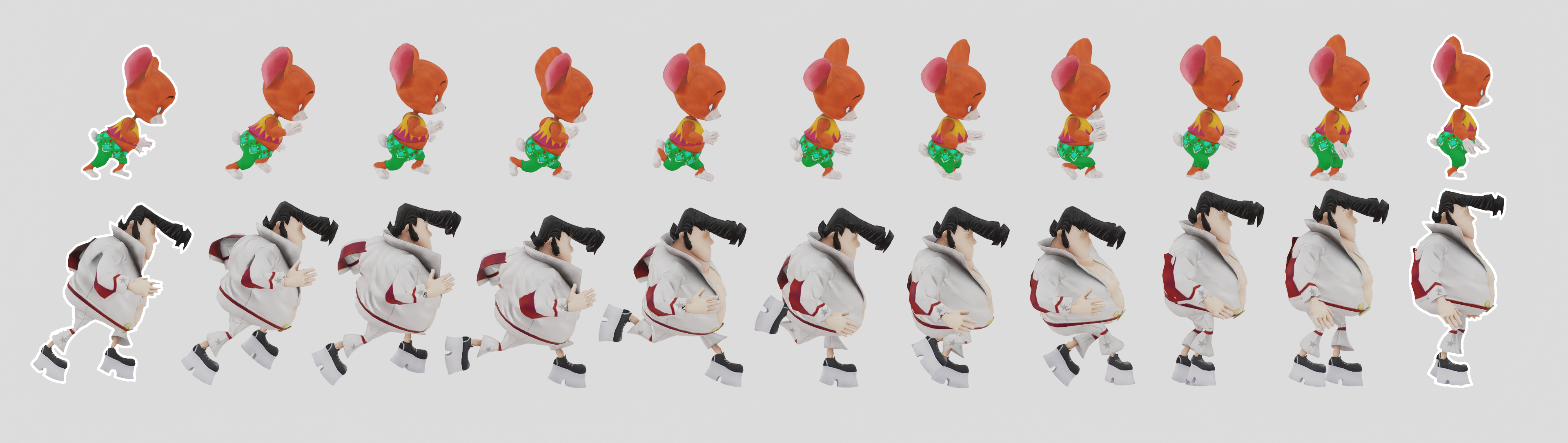}
    \caption{Qualitative results of motion in-betweening across different characters using the same keyframe constraints. Keyframes are outlined in white. Character poses are visualized every 20 frames. Stage 2 controllers are trained separately for each character.}
    \label{fig:qualitative1}
\end{figure}

\subsection{Qualitative Results}

Figure~\ref{fig:qualitative1} presents qualitative results of motion in-betweening for two distinct characters, Mouse and BigVegas, under the same keyframe constraints. Although their body proportions, skeletal structures, and mass distributions differ significantly, both characters produce coherent and high-quality transitions between the keyframes.
The in-between frames reflect natural variations and character-specific motion styles, while the keyframes, outlined in white, remain reasonably preserved and temporally aligned.

These results highlight the character-agnostic nature of our framework. In Stage 1, character-specific keyframes are first retargeted to a canonical skeleton. The diffusion model then generates a structurally consistent motion sequence from these transferred sparse keyframes. In Stage 2, a character-specific policy is trained separately for each character to adapt the motion to the physical dynamics and style of the target character. This two-stage design enables flexible in-betweening generation while maintaining motion quality and character-specific expressiveness.

\begin{figure}[tb]
    \centering
    \includegraphics[width=0.95\textwidth]{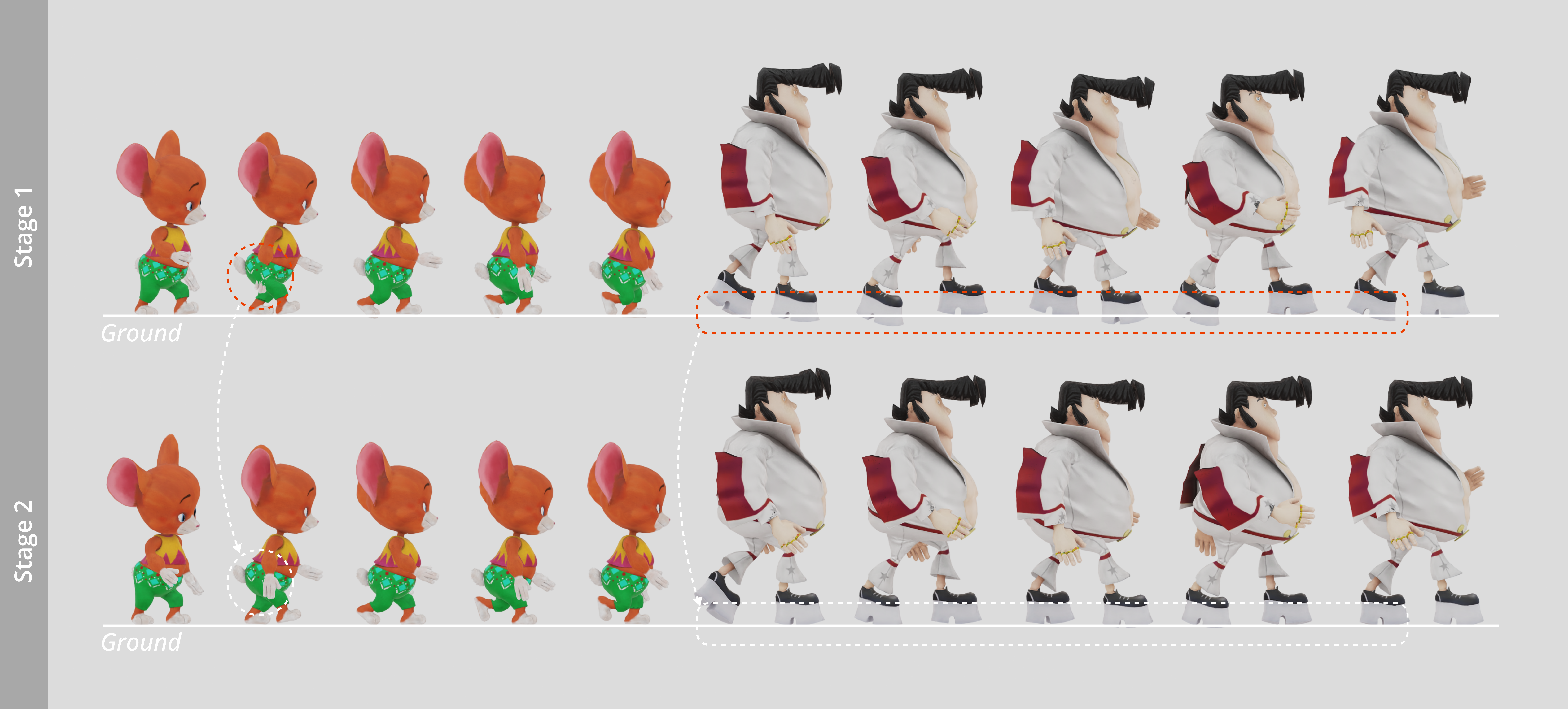}
    \caption{Comparison between Stage 1 and Stage 2 results. Stage 1 outputs are obtained by applying kinematic retargeting to the canonical motion. Stage 2 corrects artifacts from Stage 1, such as foot sliding, ground penetration, and mesh interpenetration.}
    \label{fig:qualitative2}
\end{figure}

In contrast to the generalization capability demonstrated in Figure~\ref{fig:qualitative1}, Figure~\ref{fig:qualitative2} illustrates how Stage 2 enhances motion fidelity and physical realism through physics-based adaptation. The results from Stage 1 in Figure~\ref{fig:qualitative2} are obtained by applying kinematic retargeting to the canonical motion generated in Stage 1. While this approach retains the overall structure of the motion, it often introduces noticeable artifacts, such as foot sliding, ground penetration, and mesh interpenetration. These issues are especially prominent in stylized characters such as Mouse, where exaggerated body shapes (such as a large belly) make collisions between limbs and the torso more likely.

Building on Stage 1, Stage 2 further refines the motion by enforcing physical consistency and character-specific dynamics through simulation. Character-specific controllers are trained to respect physical constraints and environmental contacts, effectively correcting foot-ground interactions and eliminating self-intersections. In addition to removing artifacts, Stage 2 enhances the stylistic expressiveness of each character. For example, BigVegas, with a heavier build and broader body, moves with a noticeable sway and slower acceleration, reflecting greater inertia and body mass. In contrast, Mouse exhibits more compact movements, with tighter upper-body motion constrained by its large torso and short limbs. These stylistic differences emerge naturally through the learning-based adaptation process, as each controller internalizes the character’s physical properties and refines the motion accordingly.

\subsection{Quantitative Results}

\subsubsection{In-Betweening with Keyframes at the Start and End}

Here we present quantitative results on the LAFAN1 dataset. All reported numbers are obtained using a simulated character with the LAFAN1 skeleton, allowing for direct comparison with existing benchmarks. This setup ensures compatibility with prior work and isolates the impact of our framework under standardized evaluation.
We first use linear interpolation between keyframes as a simple baseline. Additionally, we compare our approach with MDM\cite{tevet2022human}, which incorporates keyframes into each denoising step for diffusion-based in-betweening. DiffKFC \cite{wei2024enhanced} integrates keyframe constraints during training with a Transformer model and Dilated Mask Attention (DMA), improving keyframe adherence.

Table~\ref{tab:lafan1} reports quantitative results under the two-keyframe in-betweening setting, where only the first and last frames are provided as constraints. Across all transition lengths, our full model (Stage 1 + 2) consistently outperforms the Stage 1 baseline and prior methods in terms of K-FID, particularly for longer transitions. This improvement highlights the contribution of Stage 2 in producing more physically plausible motion. By eliminating artifacts such as foot sliding and enforcing dynamic consistency, Stage 2 introduces realistic motion details that bring the generated distribution closer to real motion data.

For K-Diversity, the differences between Stage 1 and Stage 1 + 2 are relatively small. This is expected, as the physics-based controller in Stage 2 is trained to closely track the canonical motion generated in Stage 1. As a result, the overall diversity of motion samples remains comparable, with slight variations depending on the transition length.

K-Error, which measures deviation from the keyframe positions, tends to increase slightly in Stage 1 + 2, especially for longer transitions. This is primarily due to the auto-regressive nature of the physics-based rollout in Stage 2, where small tracking errors can accumulate over time. We discuss potential strategies to address this limitation in Section~\ref{sec:discussion}.

\begin{table*}[t]
\caption{
Quantitative comparison on the LAFAN1 dataset under two-keyframe in-betweening settings. All models are trained on 200-frame clips, with only the first and last frames used as keyframes during inference. $\rightarrow$ indicates that values closer to ground truth are better.}
\label{tab:lafan1}
\centering
\small
\begin{tabular}{
l
S[table-format=1.3] S[table-format=1.3] S[table-format=1.3]|S[table-format=1.3]|
S[table-format=1.3] S[table-format=1.3] S[table-format=1.3]|S[table-format=1.3]|
S[table-format=2.3] S[table-format=2.3] S[table-format=2.3]|S[table-format=2.3]
}
\toprule
& \multicolumn{4}{c|}{K-FID $\downarrow$} & \multicolumn{4}{c|}{K-Diversity   $\rightarrow$} & \multicolumn{4}{c}{K-Error   $\downarrow$} \\
\midrule
Length                                        & \text{50}  & \text{100} & \text{200} & \text{400} & \text{50}   & \text{100} & \text{200} & \text{400} & \text{50} & \text{100} & \text{200} & \text{400}  \\ \midrule
Ground Truth                                  & \text{-}   & \text{-}   & \text{-}   & \text{-}   & 3.085       & 3.426      & 3.748      & 3.943      & \text{-}  & \text{-}   & \text{-}   & \text{-}    \\
Interpolation                                 & 6.279      & 8.106      & 9.901      & 10.987     & 0.945       & 0.861      & 0.844      & 0.854      & \text{-}  & \text{-}   & \text{-}   & \text{-}    \\
MDM {\small\cite{tevet2022human}}             & 7.600      & 7.615      & 8.156      & 22.657     & 2.645       & 2.676      & 2.265      & 1.154      & 53.835      & 57.066      & 53.505      & 169.410     \\
DiffKFC {\small\cite{wei2024enhanced}}        & 12.582     & 8.741      & 9.577      & 12.622     & 1.916       & 1.349      & 2.250      & \textbf{2.719}     & 73.797      & 57.635      & 82.940      & 105.553     \\
\midrule
Stage 1          & 4.797 & 4.895 & 5.491 & 6.109 & \textbf{2.976} & 2.798 & 2.695 & 2.411   & 10.500 & \textbf{13.376} & \textbf{17.809} & \textbf{21.682} \\ 
Stage 1 + 2      & \textbf{4.701} & \textbf{4.752} & \textbf{5.292} & \textbf{5.831} & 2.915 & \textbf{2.810} & \textbf{2.736} & 2.538   & \textbf{10.424} & 13.922 & 19.408 & 24.586 \\ 
\bottomrule
\end{tabular}
\end{table*}

\subsubsection{Uniformly-Spaced Keyframes with Fixed Length}

Table~\ref{tab:kf_density} shows results on 200-frame sequences with varying numbers of uniformly spaced keyframes. Compared to the two-keyframe setting, this experiment evaluates how the model performs under denser constraints, which can be useful for ensuring more precise control in practical applications.

Our full model consistently achieves strong K-FID scores across all intervals. While the absolute values increase as constraints become sparser, Stage 1 + 2 remains better than Stage 1 alone, confirming the effectiveness of physics-based refinement in enhancing motion realism under different supervision densities.
K-Diversity stays relatively stable across all settings. Similar to the previous experiment, the Stage 2 controller tracks the canonical motion from Stage 1 closely, leading to minimal changes in motion diversity across different keyframe intervals.
For K-Error, we again observe a slight increase after Stage 2. This pattern aligns with the expectation that auto-regressive simulation may introduce compounding deviation, especially when tracking long-range motion targets.

\begin{table*}[htbp]
\caption{
Quantitative results on LAFAN1 with varying keyframe intervals. All sequences are 200 frames long, and keyframes are uniformly spaced.}
\label{tab:kf_density}
\begin{minipage}{\textwidth}
\begin{center}
\small
\begin{tabular}{
l
S[table-format=1.3] S[table-format=1.3] S[table-format=1.3] S[table-format=1.3]|
S[table-format=1.3] S[table-format=1.3] S[table-format=1.3] S[table-format=1.3]|
S[table-format=1.3] S[table-format=2.3] S[table-format=2.3] S[table-format=2.3]
}
\toprule
& \multicolumn{4}{c|}{K-FID $\downarrow$}   & \multicolumn{4}{c|}{K-Diversity   $\rightarrow$}  & \multicolumn{4}{c}{K-Error   $\downarrow$}  \\
\midrule
Keyframe Interval & \multicolumn{1}{c}{25} & \multicolumn{1}{c}{50} & \multicolumn{1}{c}{100} & \multicolumn{1}{c|}{200} & \multicolumn{1}{c}{25} & \multicolumn{1}{c}{50} & \multicolumn{1}{c}{100} & \multicolumn{1}{c|}{200} & \multicolumn{1}{c}{25} & \multicolumn{1}{c}{50} & \multicolumn{1}{c}{100} & \multicolumn{1}{c}{200} \\
\midrule
Ground Truth                             & \text{-}           & \text{-}           & \text{-}            & \text{-}            & 3.742              & 3.745              & 3.738               & 3.748               & \text{-}           & \text{-}           & \text{-}            & \text{-} \\
MDM {\small\cite{tevet2022human}}        & 3.435              & 4.595              & 6.290               & 8.156               & 3.219              & 3.048              & 2.697               & 2.265               & 25.352             & 37.206             & 47.813              & 53.505   \\
DiffKFC {\small\cite{wei2024enhanced}}   & 1.892     & 2.878              & 4.625               & 9.577               & 3.470              & 3.350              & 3.091               & 2.250               & 19.197             & 30.751             & 49.614              & 82.940   \\ \midrule
Stage 1                                  & 1.939              & 2.472              & 3.568      & 5.491      & 3.897     & \textbf{3.592}     & 3.195      & 2.695      & \textbf{8.081}     & \textbf{10.955}    & \textbf{14.387}     & \textbf{17.809} \\  
Stage 1 + 2                              & \textbf{1.884}     & \textbf{2.406}     & \textbf{3.433}      & \textbf{5.292}      & \textbf{3.868}              & 3.574     & \textbf{3.202}      & \textbf{2.736}      & 8.105     & 11.130    & 15.007     & 19.408 \\  
\bottomrule
\end{tabular}
\end{center}
\end{minipage}
\end{table*}

\begin{table*}[htbp]
\caption{
Long-range in-betweening results on the LAFAN1 dataset. All models are trained on 200-frame clips, while inference is performed on sequences of 500 and 1000 frames with keyframes placed every 50 or 100 frames.
}
\label{tab:lafan1_long_range}
\begin{minipage}{\textwidth}
\begin{center}
\small
\begin{tabular}{
l S[table-format=1.3] S[table-format=1.3]|
  S[table-format=1.3] S[table-format=1.3]| 
  S[table-format=1.3] S[table-format=1.3]| 
  S[table-format=1.3] S[table-format=1.3]|
  S[table-format=1.3] S[table-format=1.3]| 
  S[table-format=1.3] S[table-format=1.3]}
\toprule
& \multicolumn{4}{c|}{K-FID $\downarrow$}                                                              & \multicolumn{4}{c|}{K-Diversity $\rightarrow$}                                                       & \multicolumn{4}{c}{K-Error $\downarrow$}                                                 \\
\midrule
Length            & \multicolumn{2}{c|}{500}                          & \multicolumn{2}{c|}{1000}                         & \multicolumn{2}{c|}{500}                          & \multicolumn{2}{c|}{1000}                         & \multicolumn{2}{c|}{500}                          & \multicolumn{2}{c}{1000}                         \\
\midrule
Keyframe Interval & \multicolumn{1}{c}{50} & \multicolumn{1}{c|}{100} & \multicolumn{1}{c}{50} & \multicolumn{1}{c|}{100} & \multicolumn{1}{c}{50} & \multicolumn{1}{c|}{100} & \multicolumn{1}{c}{50} & \multicolumn{1}{c|}{100} & \multicolumn{1}{c}{50} & \multicolumn{1}{c|}{100} & \multicolumn{1}{c}{50} & \multicolumn{1}{c}{100} \\
\midrule
Ground Truth      & \text{-}               & \text{-}                & \text{-}               & \text{-}                & 4.047                  & 4.053                   & 4.121               & 4.123                & \text{-}               & \text{-}          & \text{-}           & \text{-}            \\
Interpolation     & 4.478                  & 6.502                   & 3.893                  & 5.646                   & 2.626                  & 2.139                   & 2.780               & 2.386                & \text{-}               & \text{-}          & \text{-}           & \text{-}            \\
MDM \cite{tevet2022human}        & 21.145      & 22.120              & 17.458                 & 17.773                  & 3.059                  & 2.668                   & 2.537               & 2.112                & 51.091                 & 68.076            & 71.892             & 114.062             \\
DiffKFC \cite{wei2024enhanced}   & 16.010      & 15.659              & 15.066                 & 14.468                  & 4.987                  & \textbf{4.514}          & 5.509               & 5.084                & 46.783                 & 58.304            & 35.002             & 46.980              \\ \midrule
Stage 1              & 1.773         & 2.561          & 1.597         & 2.240          & \textbf{3.880}         & 3.498                   & 3.984      & 3.603       & \textbf{7.319}         & \textbf{9.739}    & \textbf{5.655}     & \textbf{7.802}      \\ 
Stage 1 + 2              & \textbf{1.722}         & \textbf{2.448}          & \textbf{1.556}         & \textbf{2.094}          & 3.857         & 3.521           & \textbf{4.011}      & \textbf{3.641}       & 7.620         & 10.211    & 6.010     & 8.749      \\ \bottomrule

\end{tabular}
\end{center}
\end{minipage}
\end{table*}

\subsubsection{Long-Range In-Betweening Results}

Table~\ref{tab:lafan1_long_range} reports results on long-range motion in-betweening tasks, with sequence lengths extended to 500 and 1000 frames. Keyframes are placed at fixed intervals (every 50 or 100 frames), and all models are trained only on 200-frame clips, making this a challenging generalization setting.

Despite being trained on significantly shorter sequences, our method generalizes well to longer motions. Stage 1 + 2 consistently achieves the best K-FID scores, demonstrating its ability to synthesize temporally coherent and physically plausible motion even over long durations. The improvement over Stage 1 is more evident as the motion length increases, suggesting that Stage 2 provides critical stability and realism during extended simulation rollouts.
K-Diversity remains high across all settings. Although Stage 2 follows the canonical motion closely, its interaction with the environment and physical constraints introduces subtle variations, which helps maintain diversity over long transitions.
K-Error increases with sequence length, as expected. While Stage 1 achieves slightly lower errors, Stage 1 + 2 remains competitive, particularly at the 1000-frame scale, where accumulated deviations are more difficult to avoid. These results indicate that our physics-based approach scales robustly to longer sequences without retraining on extended data.

\section{Discussion and Limitations}
\label{sec:discussion}

While our two-stage framework achieves good performance across a range of evaluation settings and character configurations, several limitations remain.

One notable challenge lies in maintaining keyframe consistency during long-range motion generation. In Stage 2, the physics-based controller follows the canonical motion generated in Stage 1 through auto-regressive rollout. This works reliably for short in-betweening segments, but over longer durations, small tracking errors can accumulate. These deviations, particularly in the character’s root position, can lead to noticeable misalignments at the keyframes, as reflected in the elevated K-Error observed in some cases. In practical scenarios, this issue can be alleviated in two ways. First, the motion can be divided into shorter in-betweening segments, reducing the extent of error accumulation. Second, Stage 2 can be applied iteratively. After the initial rollout, the result is adjusted by redistributing the keyframe error backward over a short temporal window. While this correction may introduce artifacts such as foot sliding, a second pass through Stage 2 can effectively remove these issues. Through a few iterations, even large deviations can be resolved.

Another limitation involves the difficulty of tracking highly dynamic or acrobatic motions, such as backflips. While our approach handles everyday movements well, it struggles with motions that fall far outside the training distribution. In such cases, the learned controller may fail to follow the canonical trajectory, resulting in noticeable discrepancies.

The use of a PD controller introduces additional challenges. Since the proportional and derivative gains are fixed hyperparameters, tuning is required for each character. Without proper tuning, the controller may exhibit delayed responses or overly damped behavior, especially during fast or complex actions. More robust control could be achieved by adopting adaptive gain schemes or replacing the PD controller with a learnable actuation model tailored to the character's dynamics.

Finally, there is an inherent limitation stemming from the gap between kinematic motion and its physical reproduction. 
While the motion generated by Stage 1 is grounded in real-world motion capture data, which naturally adheres to physical principles and contains subtle details, Stage 2’s restoration of this motion is constrained by the fidelity of the simulated character. Most physics-based character models are constructed from a small set of rigid bodies, which can only approximate the true physical structure of the character. Even with carefully designed rigging and tuned physical parameters, it is difficult to capture soft-tissue dynamics, fine-scale mass distribution, or secondary motion. Consequently, some motion details present in the original kinematic trajectory may be lost or overly smoothed. Achieving higher-fidelity adaptation would require more expressive simulation models that better reflect the physical attributes of the character.

\section{Conclusions}

We present a two-stage framework for motion in-betweening that combines the flexibility of data-driven generation with the realism of physics-based simulation. Our approach introduces two main contributions. 
First, our character-agnostic in-betweening design relaxes the traditional reliance on a fixed skeleton structure. By generating motion on a canonical skeleton in Stage 1, the framework can generalize across characters with varying skeletal structures and body proportions, enabling greater scalability and adaptability.
Second, the physics-driven motion adaptation in Stage 2 applies character-specific controllers to transform canonical motion into physically consistent and stylistically appropriate animations. This not only corrects common artifacts such as foot sliding and mesh interpenetration, but also enhances realism and conveys character-dependent motion style.
Comprehensive evaluations across qualitative and quantitative benchmarks show that our method produces high-quality motion sequences with reasonable keyframe consistency and character-specific expressiveness. We believe the proposed framework provides a practical solution for scalable and physically grounded character animation.

\bibliographystyle{unsrt}  
\bibliography{references}  

\begin{thebibliography}{10}

\bibitem{tevet2022human}
Guy Tevet, Sigal Raab, Brian Gordon, Yonatan Shafir, Daniel Cohen-Or, and Amit~H Bermano.
\newblock Human motion diffusion model.
\newblock In {\em The Eleventh International Conference on Learning Representations (ICLR)}, 2023.

\bibitem{qin2024robust}
Jia Qin, Peng Yan, and Bo~An.
\newblock {Robust Diffusion-based Motion In-betweening}.
\newblock {\em Computer Graphics Forum}, 43(7):e15260, 2024.

\bibitem{qin2022motion}
Jia Qin, Youyi Zheng, and Kun Zhou.
\newblock Motion in-betweening via two-stage transformers.
\newblock {\em ACM Transactions on Graphics (TOG)}, 41(6):184--1, 2022.

\bibitem{oreshkin2023motion}
Boris~N Oreshkin, Antonios Valkanas, F{\'e}lix~G Harvey, Louis-Simon M{\'e}nard, Florent Bocquelet, and Mark~J Coates.
\newblock Motion in-betweening via deep {$\Delta$}-interpolator.
\newblock {\em IEEE Transactions on Visualization and Computer Graphics}, 2023.

\bibitem{witkin1988spacetime}
Andrew Witkin and Michael Kass.
\newblock Spacetime constraints.
\newblock {\em ACM Siggraph Computer Graphics}, 22(4):159--168, 1988.

\bibitem{liu2005learning}
C~Karen Liu, Aaron Hertzmann, and Zoran Popovi{\'c}.
\newblock Learning physics-based motion style with nonlinear inverse optimization.
\newblock {\em ACM Transactions on Graphics (TOG)}, 24(3):1071--1081, 2005.

\bibitem{peng2018deepmimic}
Xue~Bin Peng, Pieter Abbeel, Sergey Levine, and Michiel Van~de Panne.
\newblock Deepmimic: Example-guided deep reinforcement learning of physics-based character skills.
\newblock {\em ACM Transactions On Graphics (TOG)}, 37(4):1--14, 2018.

\bibitem{winkler2022questsim}
Alexander Winkler, Jungdam Won, and Yuting Ye.
\newblock Questsim: Human motion tracking from sparse sensors with simulated avatars.
\newblock In {\em SIGGRAPH Asia 2022 Conference Papers}, pages 1--8, 2022.

\bibitem{reda2023physics}
Daniele Reda, Jungdam Won, Yuting Ye, Michiel van~de Panne, and Alexander Winkler.
\newblock Physics-based motion retargeting from sparse inputs.
\newblock {\em Proceedings of the ACM on Computer Graphics and Interactive Techniques}, 6(3):1--19, 2023.

\bibitem{tessler2024maskedmimic}
Chen Tessler, Yunrong Guo, Ofir Nabati, Gal Chechik, and Xue~Bin Peng.
\newblock Maskedmimic: Unified physics-based character control through masked motion inpainting.
\newblock {\em ACM Transactions on Graphics (TOG)}, 43(6):1--21, 2024.

\bibitem{fussell2021supertrack}
Levi Fussell, Kevin Bergamin, and Daniel Holden.
\newblock Supertrack: Motion tracking for physically simulated characters using supervised learning.
\newblock {\em ACM Transactions on Graphics (TOG)}, 40(6):1--13, 2021.

\bibitem{yuan2023physdiff}
Ye~Yuan, Jiaming Song, Umar Iqbal, Arash Vahdat, and Jan Kautz.
\newblock Physdiff: Physics-guided human motion diffusion model.
\newblock In {\em Proceedings of the IEEE/CVF International Conference on Computer Vision}, pages 16010--16021, 2023.

\bibitem{ramesh2022hierarchical}
Aditya Ramesh, Prafulla Dhariwal, Alex Nichol, Casey Chu, and Mark Chen.
\newblock Hierarchical text-conditional image generation with clip latents.
\newblock {\em arXiv preprint arXiv:2204.06125}, 1(2):3, 2022.

\bibitem{dhariwal2021diffusion}
Prafulla Dhariwal and Alexander Nichol.
\newblock Diffusion models beat gans on image synthesis.
\newblock {\em Advances in neural information processing systems}, 34:8780--8794, 2021.

\bibitem{rombach2022high}
Robin Rombach, Andreas Blattmann, Dominik Lorenz, Patrick Esser, and Bj{\"o}rn Ommer.
\newblock High-resolution image synthesis with latent diffusion models.
\newblock In {\em Proceedings of the IEEE/CVF conference on computer vision and pattern recognition}, pages 10684--10695, 2022.

\bibitem{saharia2022photorealistic}
Chitwan Saharia, William Chan, Saurabh Saxena, Lala Li, Jay Whang, Emily~L Denton, Kamyar Ghasemipour, Raphael Gontijo~Lopes, Burcu Karagol~Ayan, Tim Salimans, et~al.
\newblock Photorealistic text-to-image diffusion models with deep language understanding.
\newblock {\em Advances in neural information processing systems}, 35:36479--36494, 2022.

\bibitem{ho2022imagen}
Jonathan Ho, William Chan, Chitwan Saharia, Jay Whang, Ruiqi Gao, Alexey Gritsenko, Diederik~P Kingma, Ben Poole, Mohammad Norouzi, David~J Fleet, et~al.
\newblock Imagen video: High definition video generation with diffusion models.
\newblock {\em arXiv preprint arXiv:2210.02303}, 2022.

\bibitem{ho2022video}
Jonathan Ho, Tim Salimans, Alexey Gritsenko, William Chan, Mohammad Norouzi, and David~J Fleet.
\newblock Video diffusion models.
\newblock {\em Advances in Neural Information Processing Systems}, 35:8633--8646, 2022.

\bibitem{luo2021diffusion}
Shitong Luo and Wei Hu.
\newblock Diffusion probabilistic models for 3d point cloud generation.
\newblock In {\em Proceedings of the IEEE/CVF Conference on Computer Vision and Pattern Recognition}, pages 2837--2845, 2021.

\bibitem{zhou20213d}
Linqi Zhou, Yilun Du, and Jiajun Wu.
\newblock 3d shape generation and completion through point-voxel diffusion.
\newblock In {\em Proceedings of the IEEE/CVF international conference on computer vision}, pages 5826--5835, 2021.

\bibitem{kong2020diffwave}
Zhifeng Kong, Wei Ping, Jiaji Huang, Kexin Zhao, and Bryan Catanzaro.
\newblock Diffwave: A versatile diffusion model for audio synthesis.
\newblock {\em arXiv preprint arXiv:2009.09761}, 2020.

\bibitem{song2020score}
Yang Song, Jascha Sohl-Dickstein, Diederik~P Kingma, Abhishek Kumar, Stefano Ermon, and Ben Poole.
\newblock Score-based generative modeling through stochastic differential equations.
\newblock {\em arXiv preprint arXiv:2011.13456}, 2020.

\bibitem{ho2020denoising}
Jonathan Ho, Ajay Jain, and Pieter Abbeel.
\newblock Denoising diffusion probabilistic models.
\newblock {\em Advances in neural information processing systems}, 33:6840--6851, 2020.

\bibitem{ho2022classifier}
Jonathan Ho and Tim Salimans.
\newblock Classifier-free diffusion guidance.
\newblock {\em arXiv preprint arXiv:2207.12598}, 2022.

\bibitem{lipman2022flow}
Yaron Lipman, Ricky~TQ Chen, Heli Ben-Hamu, Maximilian Nickel, and Matt Le.
\newblock Flow matching for generative modeling.
\newblock {\em arXiv preprint arXiv:2210.02747}, 2022.

\bibitem{poole2022dreamfusion}
Ben Poole, Ajay Jain, Jonathan~T Barron, and Ben Mildenhall.
\newblock Dreamfusion: Text-to-3d using 2d diffusion.
\newblock {\em arXiv preprint arXiv:2209.14988}, 2022.

\bibitem{song2020denoising}
Jiaming Song, Chenlin Meng, and Stefano Ermon.
\newblock Denoising diffusion implicit models.
\newblock {\em arXiv preprint arXiv:2010.02502}, 2020.

\bibitem{dabral2023mofusion}
Rishabh Dabral, Muhammad~Hamza Mughal, Vladislav Golyanik, and Christian Theobalt.
\newblock Mofusion: A framework for denoising-diffusion-based motion synthesis.
\newblock In {\em Proceedings of the IEEE/CVF conference on computer vision and pattern recognition}, pages 9760--9770, 2023.

\bibitem{ma2022pretrained}
Jianxin Ma, Shuai Bai, and Chang Zhou.
\newblock Pretrained diffusion models for unified human motion synthesis.
\newblock {\em arXiv preprint arXiv:2212.02837}, 2022.

\bibitem{zhou2023ude}
Zixiang Zhou and Baoyuan Wang.
\newblock Ude: A unified driving engine for human motion generation.
\newblock In {\em Proceedings of the IEEE/CVF Conference on Computer Vision and Pattern Recognition}, pages 5632--5641, 2023.

\bibitem{kim2023flame}
Jihoon Kim, Jiseob Kim, and Sungjoon Choi.
\newblock Flame: Free-form language-based motion synthesis \& editing.
\newblock In {\em Proceedings of the AAAI Conference on Artificial Intelligence}, volume~37, pages 8255--8263, 2023.

\bibitem{chen2024taming}
Rui Chen, Mingyi Shi, Shaoli Huang, Ping Tan, Taku Komura, and Xuelin Chen.
\newblock Taming diffusion probabilistic models for character control.
\newblock In {\em ACM SIGGRAPH 2024 Conference Papers}, pages 1--10, 2024.

\bibitem{zhang2024motiondiffuse}
Mingyuan Zhang, Zhongang Cai, Liang Pan, Fangzhou Hong, Xinying Guo, Lei Yang, and Ziwei Liu.
\newblock Motiondiffuse: Text-driven human motion generation with diffusion model.
\newblock {\em IEEE Transactions on Pattern Analysis and Machine Intelligence}, 2024.

\bibitem{chen2023executing}
Xin Chen, Biao Jiang, Wen Liu, Zilong Huang, Bin Fu, Tao Chen, and Gang Yu.
\newblock Executing your commands via motion diffusion in latent space.
\newblock In {\em Proceedings of the IEEE/CVF Conference on Computer Vision and Pattern Recognition}, pages 18000--18010, 2023.

\bibitem{shafir2024human}
Yonatan Shafir, Guy Tevet, Roy Kapon, and Amit~H Bermano.
\newblock Human motion diffusion model.
\newblock In {\em The Twelfth International Conference on Learning Representations (ICLR)}, 2024.

\bibitem{ciccone2019tangent}
Lo{\"\i}c Ciccone, Cengiz {\"O}ztireli, and Robert~W Sumner.
\newblock Tangent-space optimization for interactive animation control.
\newblock {\em ACM Transactions on Graphics (TOG)}, 38(4):1--10, 2019.

\bibitem{ngo1993spacetime}
J~Thomas Ngo and Joe Marks.
\newblock Spacetime constraints revisited.
\newblock In {\em Proceedings of the 20th annual conference on Computer graphics and interactive techniques}, pages 343--350, 1993.

\bibitem{witkin1995motion}
Andrew Witkin and Zoran Popovic.
\newblock Motion warping.
\newblock In {\em Proceedings of the 22nd annual conference on Computer graphics and interactive techniques}, pages 105--108, 1995.

\bibitem{gleicher1997motion}
Michael Gleicher.
\newblock Motion editing with spacetime constraints.
\newblock In {\em Proceedings of the 1997 symposium on Interactive 3D graphics}, pages 139--ff, 1997.

\bibitem{chai2007constraint}
Jinxiang Chai and Jessica~K. Hodgins.
\newblock Constraint-based motion optimization using a statistical dynamic model.
\newblock {\em ACM Transactions on Graphics (TOG)}, 26(3):8–es, jul 2007.

\bibitem{min2009interactive}
Jianyuan Min, Yen-Lin Chen, and Jinxiang Chai.
\newblock Interactive generation of human animation with deformable motion models.
\newblock {\em ACM Transactions on Graphics (TOG)}, 29(1):1--12, 2009.

\bibitem{wang2007gaussian}
Jack~M Wang, David~J Fleet, and Aaron Hertzmann.
\newblock Gaussian process dynamical models for human motion.
\newblock {\em IEEE transactions on pattern analysis and machine intelligence}, 30(2):283--298, 2007.

\bibitem{lehrmann2014efficient}
Andreas~M Lehrmann, Peter~V Gehler, and Sebastian Nowozin.
\newblock Efficient nonlinear markov models for human motion.
\newblock In {\em Proceedings of the IEEE conference on computer vision and pattern recognition}, pages 1314--1321, 2014.

\bibitem{harvey2018recurrent}
F{\'e}lix~G Harvey and Christopher Pal.
\newblock Recurrent transition networks for character locomotion.
\newblock In {\em SIGGRAPH Asia 2018 Technical Briefs}, pages 1--4, 2018.

\bibitem{harvey2020robust}
F{\'e}lix~G Harvey, Mike Yurick, Derek Nowrouzezahrai, and Christopher Pal.
\newblock Robust motion in-betweening.
\newblock {\em ACM Transactions on Graphics (TOG)}, 39(4):60--1, 2020.

\bibitem{hernandez2019human}
Alejandro Hernandez, Jurgen Gall, and Francesc Moreno-Noguer.
\newblock Human motion prediction via spatio-temporal inpainting.
\newblock In {\em Proceedings of the IEEE/CVF International Conference on Computer Vision}, pages 7134--7143, 2019.

\bibitem{kaufmann2020convolutional}
Manuel Kaufmann, Emre Aksan, Jie Song, Fabrizio Pece, Remo Ziegler, and Otmar Hilliges.
\newblock Convolutional autoencoders for human motion infilling.
\newblock In {\em 2020 International Conference on 3D Vision (3DV)}, pages 918--927. IEEE, 2020.

\bibitem{duan2021single}
Yinglin Duan, Tianyang Shi, Zhengxia Zou, Yenan Lin, Zhehui Qian, Bohan Zhang, and Yi~Yuan.
\newblock Single-shot motion completion with transformer.
\newblock {\em arXiv preprint arXiv:2103.00776}, 2021.

\bibitem{karunratanakul2023guided}
Korrawe Karunratanakul, Konpat Preechakul, Supasorn Suwajanakorn, and Siyu Tang.
\newblock Guided motion diffusion for controllable human motion synthesis.
\newblock In {\em Proceedings of the IEEE/CVF International Conference on Computer Vision}, pages 2151--2162, 2023.

\bibitem{wei2024enhanced}
Dong Wei, Xiaoning Sun, Huaijiang Sun, Shengxiang Hu, Bin Li, Weiqing Li, and Jianfeng Lu.
\newblock Enhanced fine-grained motion diffusion for text-driven human motion synthesis.
\newblock In {\em Proceedings of the AAAI Conference on Artificial Intelligence}, volume~38, pages 5876--5884, 2024.

\bibitem{zhou2019continuity}
Yi~Zhou, Connelly Barnes, Jingwan Lu, Jimei Yang, and Hao Li.
\newblock On the continuity of rotation representations in neural networks.
\newblock In {\em Proceedings of the IEEE/CVF conference on computer vision and pattern recognition}, pages 5745--5753, 2019.

\bibitem{huang2018music}
Cheng-Zhi~Anna Huang, Ashish Vaswani, Jakob Uszkoreit, Noam Shazeer, Ian Simon, Curtis Hawthorne, Andrew~M Dai, Matthew~D Hoffman, Monica Dinculescu, and Douglas Eck.
\newblock Music transformer.
\newblock {\em arXiv preprint arXiv:1809.04281}, 2018.

\bibitem{mixamo}
{Adobe Inc.}
\newblock Mixamo.
\newblock \url{https://www.mixamo.com}, 2025.

\bibitem{lee2023questenvsim}
Sunmin Lee, Sebastian Starke, Yuting Ye, Jungdam Won, and Alexander Winkler.
\newblock Questenvsim: Environment-aware simulated motion tracking from sparse sensors.
\newblock In {\em ACM SIGGRAPH 2023 Conference Proceedings}, pages 1--9, 2023.

\bibitem{paszke2019pytorch}
A~Paszke.
\newblock Pytorch: An imperative style, high-performance deep learning library.
\newblock {\em arXiv preprint arXiv:1912.01703}, 2019.

\bibitem{schulman2017proximal}
John Schulman, Filip Wolski, Prafulla Dhariwal, Alec Radford, and Oleg Klimov.
\newblock Proximal policy optimization algorithms.
\newblock {\em arXiv preprint arXiv:1707.06347}, 2017.

\bibitem{kingma2014adam}
Diederik~P Kingma and Jimmy Ba.
\newblock Adam: A method for stochastic optimization.
\newblock {\em arXiv preprint arXiv:1412.6980}, 2014.

\bibitem{mittal2023orbit}
Mayank Mittal, Calvin Yu, Qinxi Yu, Jingzhou Liu, Nikita Rudin, David Hoeller, Jia~Lin Yuan, Ritvik Singh, Yunrong Guo, Hammad Mazhar, Ajay Mandlekar, Buck Babich, Gavriel State, Marco Hutter, and Animesh Garg.
\newblock Orbit: A unified simulation framework for interactive robot learning environments.
\newblock {\em IEEE Robotics and Automation Letters}, 8(6):3740--3747, 2023.

\end{thebibliography}






\end{document}